# New trends in mechanistic transdermal drug delivery modelling: Towards an accurate description of skin microstructure

Daniel Sebastia-Saez[1,*], Adam Burbidge[2], Jan Engmann[2], Marco Ramaioli[1]

[1]*Department of Chemical and Process Engineering, University of Surrey, GU2 7XH Guildford, United Kingdom.*

[2]*Nestlé Research Center, Route de Jorat 57, vers-chez-les Blanc, 1000 Lausanne 26, Switzerland*

## Table of Contents

**Abstract**

Interest for *in silico* modelling of the absorption of xenobiotics into the skin has been growing in the last years, owing to their lower cost compared to experimental alternatives, and the desire to avoid animal experimentation. This review presents an overview of Physiologically-Based Pharmacokinetic (PBPK) models and focuses on recent, modelling approaches, such as Finite Element and Lattice Boltzmann. These methods allow for a detailed geometric representation of the skin microstructure, in contrast to classic QSPR and compartmental models. Morphological features of the skin such as the bricks and mortar description of the stratum corneum, hair follicles, and the pilosebaceous unit can therefore be represented more accurately, allowing a better description of the interaction of cosmetics with the skin. This review also highlights several perspectives to further develop these models in directions relevant to industry.

*Corresponding author: Dr Daniel Sebastia-Saez. E-mail address: j.sebastiasaez@surrey.ac.uk
Tel. +44 (0) 7719 715 575

# 1. Introduction.

Maintaining healthy skin is an important aspect of maintaining health of the organism as a whole [1], [2]. As such, the skin has received considerable attention from the cosmetic and pharmaceutical industry for a long time. Human skin exposure to chemicals can be unintentional (industrial hazards derived from chemicals, pollutants, and cleaning products) [3], or intentional (cosmetics or medical drugs of topical application) [4], [5]. Topical administration of cosmetic products is a difficult engineering task which has been the subject of numerous studies, both numerical and experimental [6], [7]. The skin outer layer (the stratum corneum), which is comprised of keratinised cells and lipids, acts as a natural barrier against external agents and therefore, makes drug and cosmetic topical absorption a considerable challenge [8]. On the other hand, topical administration of drugs and cosmetics allows the active material to be provided in a locally-targeted fashion, hence the interest of the pharmaceutical industry in developing ever more descriptive models. Avoiding animal experimentation plays also a pivotal role in the recent interest in detailed in-silico modelling of the skin microstructure [9].

Transdermal drug delivery (TDD) has the potential to overcome many limitations posed by other drug administration methods such as the oral and the parenteral route [10], [11], which limit the possibility of administration of some peptides and proteins because of size restrictions (oral) or require trained personnel among other drawbacks (parenteral). There are however significant challenges to optimising molecular transport across the skin, and new modelling approaches making use of algorithms such as the finite element method or molecular dynamics, which allows representation of skin microstructural features, will surely play an increased role in the near future [12]–[15]. Solving flow and diffusion equations in a discretised representation of the skin microstructure has encouraged research efforts towards complementing and enhancing the amount of information provided by QSPR and compartmental models. As a result, bio-engineered skin substitutes and *in silico* analyses are pointed out as potential substitutes for animal models in recent times [16]. This review focusses on the evolution of *in silico* models and highlights the need for fully detailed discretized-microstructure (DM from now on) physiologically-based pharmacokinetic (PBPK) models instead of modelling methods such as Quantitative Structure-Property Relationships (QSPR) and compartmental models. DM-PBPK models also prove a valuable tool to complement *in vitro* assays [17]. This is the context of this review.

The review starts with a description of the QSPR and compartmental models—the first modelling categories for skin permeability that appeared in the literature—discussing their capabilities and limitations. The article subsequently focuses on highlighting the improvement and further capabilities in DM-PBPK models since these were first reported in a review published by Mitragotri et al. [18] in 2011 and by de Monte et al. in 2015 [19], who gathered and discussed the first FEM models describing the microstructure of the skin main barrier, i.e. the stratum corneum (Chandrasekaran et al., 1978 [20]; Rim et al., 2005 [15], Frasch and Barbero, 2003 [21]). Models developed recently have focussed on the description of the geometric and mechanical properties of the skin appendages, such as hair follicles, which play an important role on the diffusion of drugs and cosmetics. Following this line, this review finishes with a general summary of future perspectives suggested by industrial partners at Nestlé Research Center (Vevey, Switzerland) to exploit the capabilities of DM-PBPK models.

## 2. An overview of classic modelling categories and their limitations: QSPR and compartmental models

QSPR models are regression models frequently used in chemistry and biology to determine hazards posed by substances. Being based on regression of data obtained experimentally from *in vitro* assays at steady-state conditions, QSPR models are restricted to the application being considered and cannot be extrapolated to other conditions or used to describe microstructural features such as skin appendages. QSPR models aim on the other hand to obtain accurate predictions of the permeability of the skin to a given substance, and thus constitute a useful tool for risk assessment. The model provided by Potts and Guy in 1992 [22] has been recently used to evaluate the extension of the Threshold of Toxicological Concern (TTC) guidelines issued by the European Union ([23]; Potts and Guy, 1992 and 1995 [22], [24]). They have also been used in the frame of the European Union COSMOS project, which aims to study the toxicity of cosmetic products. The input to QSPR models is the solubility in lipids, the partition coefficients and the molecular weight. Measuring solubility in lipids presents some experimental difficulties however and surrogates such as octanol are often sought. The partition coefficients between the vehicle and the lipids are necessary to determine the concentration of the xenobiotic component at the outermost part of the stratum corneum. Once the permeability is established, the next step in the development of the model is the application of the diffusion laws to obtain the mass transfer rate, total amount of substance being absorbed, et cetera, thus assessing the penetration potential of a given substance and the related hazards. The solute transport through the skin under steady-state conditions can be calculated by integrating Fick's first law, which allows for the calculation of the maximum flux. The integrated expression for the first Fick's law gives the total amount of substance absorbed by the skin as a function of time and depth into the skin. Immediately after the application of the topical solution there is a lag time which varies considerably with the solute size—from 0.5 h for the permeation of water through the stratum corneum to hundreds or thousands of hours for larger solutes.

*Table 1 Summary of available databases with permeability coefficients available for QSPR models published in the scientific literature as of 2019.*

| Author | No. of partition coeff. values | Range of Molecular weight | $log(K_{ow})$ ($K_{ow}$ being the octanol- | Further comments |
|---|---|---|---|---|

|  |  | [g/mol] | water partition coeff.) |  |
| --- | --- | --- | --- | --- |
| Brown et al. (2016) [25] | 392 from 245 compounds | 18–765 | -6.8 to 7.6 | Only from human skin and using water as the vehicle. |
| Alves et al. (2015) [26] | - | - | -4.85 to -0.94 | Data from human and rodent skin samples. 185 and 96 compounds considered, respectively. |
| Chen et al. (2013) [27] | 71 from 35 compounds | - | - | Only from human skin and hydrophilic chemicals. |
| Wilschut et al. (1995) [28] | 123 from 99 compounds | - | - | Various chemical classes considered including monoaromatic hydrocarbons, volatile halogenated hydrocarbons, phenols and steroids. Useful for industrial hazard evaluation, but lacking components used in pharmaceutical and cosmeceutical industries. |
| Patel et al. (2002) [29] | 186 from 158 compounds | - | - | High diversity of compounds structure. |
| Vecchia and Bunge (2003) [30] | 170 from 127 compounds | 18–584 | -3.1 to 4.6 | - |
| Magnusson et al. (2004) [31] | - | 18–765 | -5.7 to 8.7 | The database directly provides the values of the maximum flux. |
| EDETOX Database [32] | 4800 from 320 compounds | - | - | Considerable additional information (description of the skin samples, lag time, exposure time, etc. |

The value of the maximum flux should be corrected depending on the characteristics of a given xenobiotic. The European Commission has issued recommended values of the percentage of absorption rate to set guidelines for risk assessment policies. A value of 10% is used in case of solutes with a molecular weight greater than 500 Da and octanol-water partition coefficient smaller than -1 or higher than 4. Otherwise 100% of the maximum flux applies [33]. Considering that the vehicle can be approximated to an infinite source, the maximum flux is the result of multiplying the permeability coefficient and the solubility of the xenobiotic.

QSPR models have been reported to obtain accurate calculations of the permeability of different substances via *in vitro* experimentation, being a widespread assumption that the stratum corneum behaves as a homogeneous layer. Details on the morphology of the stratum corneum are neglected, and thus the quantification of the relative importance between the three different absorption routes within the stratum corneum (intracellular, lipidic, and appendageal) cannot be assessed. Another common assumption is that the resistance to skin absorption comes only from the stratum corneum, neglecting the resistance posed by the viable epidermis and upper avascular dermis, and possibly resulting in an over-prediction of the solute concentration reached in the highly vascularised layers of the skin.

A comprehensive set of experimental data on permeation coefficients of hazardous substances has been gathered in data bases, although they present significant variations owed to factors concerning the conditions of the experiment. Some of these data bases published in the literature in the last three decades are summarised in Table 1.

QSPR models cannot be extrapolated to conditions other than those upon which the data have been obtained, and therefore their use as predictive tools with the potential of saving money and resources for experimentation remains limited. QSPR models are still needed though to support mechanistic models by providing the values of the parameters needed to discretize and solve the diffusion equations.

As for the quality of these models to represent the penetration of xenobiotic substances within the skin, some QSPR models present a high degree of variability reflected in the poor correlation coefficients. The simplest QSPR models obtain the permeation coefficient from the linear regression of basic structural descriptors, namely the molecular size (denoted by the molecular volume or the molecular weight) and the hydrophobicity (expressed by the octanol-water partition coefficient). Other QSPR models utilize complex parameters such as quantum chemical indices.

*Table 2 Summary of QSPR models based on linear regression and the quality of the fit*

| Author | Correlation coefficient | Comment |
| --- | --- | --- |
| Potts and Guy (1992) [22] | 0.670 | Hydrophobicity and molecule size as structural descriptors |
| Moss and Cronin (2002) [34] | 0.820 | Id. |
| Magnusson et al (2004) [31] | 0.847 | Id. |
| Basak et al. (2007) [35] | 0.67–0.87 | Model including not only two structural descriptors but also shape descriptors and quantum chemical indices. |

Table 2 presents a summary of several QSPR models with their correlation coefficients in order to measure the goodness of the fitting. The models of and Guy (1992, 1995) [22], [24], Moss and Cronin (2002) [34], and Magnusson et al. (2004) [31] are obtained using only the hydrophobicity and the molecular size, and present correlation coefficients which range from 0.670 to 0.847. The model of Basak et al. (2007) [35] considers more than two descriptors, including more complex molecular aspects. The goodness of the fitting does not necessarily improve with respect to the other models though. To prevent prediction errors, this variability needs to be considered when developing mechanistic models which use QSPR as a source of diffusion-related data.

The complexity of QSPR models is still growing nowadays, and are being used in complex aspects concerning the design of new drugs [36].

QSPR models have been extensively used in pharmacology to predict skin permeability of chemicals through the different layers which form the skin. In their most basic configuration, QSPR are mathematical one-dimensional models where data such as molecular weights, octanol-water partition coefficients and solubility in water are input to assess the permeability of a given compound. The one-dimensional character of QSPR models gives way to a lack of detail on the visualisation of the phenomena at hand. Also, QSPR models rely on experimental data obtained previously, which limits their versatility. QSPR models have been extensively used in the past for identification of potential hazards and risk assessment [37].

On the other hand, mechanistic models called physiologically-based pharmacokinetic (PBPK) models include lumped-parameter models, i.e. the compartmental approach, and the recently-developed discretised-microstructure (DM) PBPK approach (Lu et al., 2012), which to date offers the greatest degree of detail among the modelling categories used to describe pharmacokinetic systems. The compartmental approach adds complexity to previously-discussed QSPR as they offer the possibility of obtaining variations in time and space of the concentration of a particular species being diffused into the skin layers [38]. In a compartmental model, each skin layer is represented by a single compartment with no distinct internal features and constant species concentration values where an intertwined system of linearised diffusion equations is solved [39]:

$$\frac{dC_1}{dt} = k(K_{SC}C_v + C_2 - 2C_1), \ldots \quad (1)$$

$$\ldots\frac{dC_i}{dt} = k(C_{i-1} + C_{i+1} - 2C_i), \ldots$$

$$\ldots\frac{dC_n}{dt} = k(C_{n-1} - 2C_n).$$

Compartmental models offer a fair trade between computational economy and purpose. Simple compartmental models were limited to the study of the mass exchange between two compartments, namely the low permeability stratum corneum, and a high permeability second compartment which represents the viable epidermis and the upper avascular dermis [40]. More complex models, featuring up to five compartments in a cascade disposition have been reported in the literature in order to get a more accurate prediction of xenobiotic concentration at the target layer [41]. Generalisation to any number of compartments was introduced by Anissimov et al. [39].

Compartmental models give a single value of the xenobiotic concentration for each compartment as a result, i.e. one compartment per skin layer. They can thus be considered as an approximate attempt to provide a description of the concentration-depth profiles caused by the absorption of a xenobiotic component within the skin. In the limit, an infinite number of compartments would provide a smooth concentration-depth profile. Compartmental models present thus an approximation to the concentration-depth profiles and focus on the calculation of the mass transfer rate between layers, contrarily to previously discussed QSPR models, the objective of which is the calculation of permeability by fitting experimental data [42]. Extra compartments can be added to these models in order to account for chemical binding and volatilization of the component after topical application [43], but still lack details on the interaction between the skin microstructure and the active component [44], [45].

3. **Discretised-microstructure Physiologically-based Pharmacokinetic (DM-PBPK) models.**

DM-PBPK models have been developed to gain more detail in the description of skin pharmacokinetics. A DM-PBPK modelling approach consists in reproducing the actual microstructure of the skin through a Computer Aided Design (CAD) software and discretise it afterwards. The diffusion equations, or a combination of the latter and the Navier-Stokes equation if there is flow of blood involved, can be solved in the discretised domain to provide a more realistic prediction of the absorption process. DM-PBPK models still rely however on experimental data of the diffusion and partition coefficients, and on an appropriate representation of the skin microstructure. These models are therefore useful to provide insight on some specifics of the problem such as detailed concentration-depth profiles, which would otherwise be difficult to get using experimental techniques. The above mentioned categories are not exclusive however, and can be combined as in the case of Savoca et al. (2018) [46] and Abbiati et al. [47], [48] who combined a compartmental set-up with a discretized solution of the diffusion equations.

Upon application on the skin, permeation occurs by passive diffusion through the skin layers and its appendages, e.g. hair follicles and pilosebaceous glands [49]. Therefore, a comprehensive description of these layers and appendages is necessary to accomplish detailed modelling of its microstructure. Also, the hair follicle can constitute a shortcut to skip the first barrier (the stratum corneum), enabling transfer into the skin. The hair follicle has therefore received considerable attention lately in terms of research ([50]; [51]; [52]).

The skin is comprised of three main layers (the epidermis, the dermis and the hypodermis). The epidermis lies in contact with the open air and provides waterproof barrier as well as pigmentation. Two layers can be clearly differentiated in the epidermis: the stratum corneum, which constitutes the actual barrier given its low permeability, and the viable epidermis. The dermis is a complex combination of blood vessels, hair follicles, sebaceous glands, fibroblasts and nerves. Collagen and elastin, substances that are crucial to a healthy skin, are synthesised in the dermis, too. The rich vascularization of the dermis can often induce an undesirable distribution of topically applied products to the rest of the organism via the circulatory system. The viable epidermis and dermis can be simplified in terms of modelling as homogeneous layers composed of approximately 40% water, 40% proteins, and 20% lipids ([53]; [54]). Most research efforts focus on the stratum corneum though, which is the outermost layer of the epidermis and constitutes an effective barrier against penetration of alien substances given its morphology. The stratum corneum is formed of two phases with different effects on the mass absorption rate: one formed mainly by keratinocytes and the other by lipids. Most of the mass transfer occurs through the lipidic route given its greater permeability relative to the keratinocyte route. Figure 1 shows a schematic of the stratum corneum used in a DM-PBPK simulation [55]. The latter provides more detail on how the xenobiotic crosses the stratum corneum than compartmental models, which provide a general idea of the diffusion process by averaging properties of the constituent components of the stratum corneum. The hypodermis consists of adipose and connective tissue and is also highly vascularized. DM-PBPK models can therefore provide a detailed geometric representation of these layers including any irregularity in its microstructure and all the physical properties associated with them. For this reason, recent research trends lean towards using DM-PBPK models, although QSPR and compartmental models are still widespread.

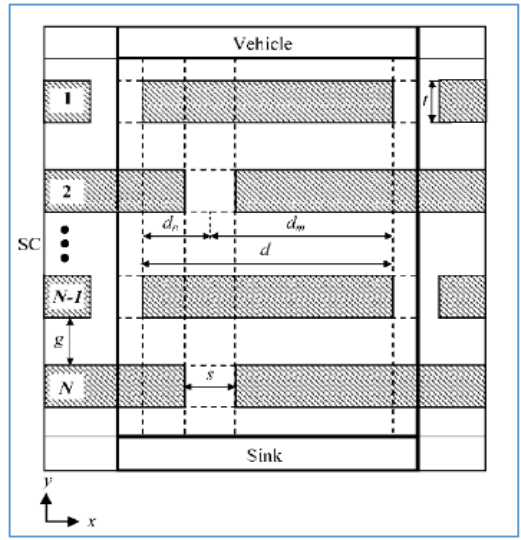

*Figure 1 Schematic plot of the stratum corneum. DM-PBPK models can include high fidelity representations of the stratum corneum microstructure. Reprinted (adapted) with permission from (Chen et al. Ind. Eng. Chem. Res. 47(17), pp. 6465–6472). Copyright (2008) American Chemical Society.*

Figure 2 shows visually the difference between compartmental and DM-PBPK models. Figure 2a shows a schematic of the compartmental model used by Anissimov et al. (2013) [39], where each box represents one of the skin layers. The microstructure of those layers is omitted and only the relationship between them is represented. Figure 2b on the other hand shows the representation of the hair follicle in the DM-PBPK model presented by Kattou et al. (2017) [56], where each skin layer has a finite thickness and is defined by its physical properties. The latter enables computing detailed time-dependent concentration profiles as the one represented in Figure 2c.

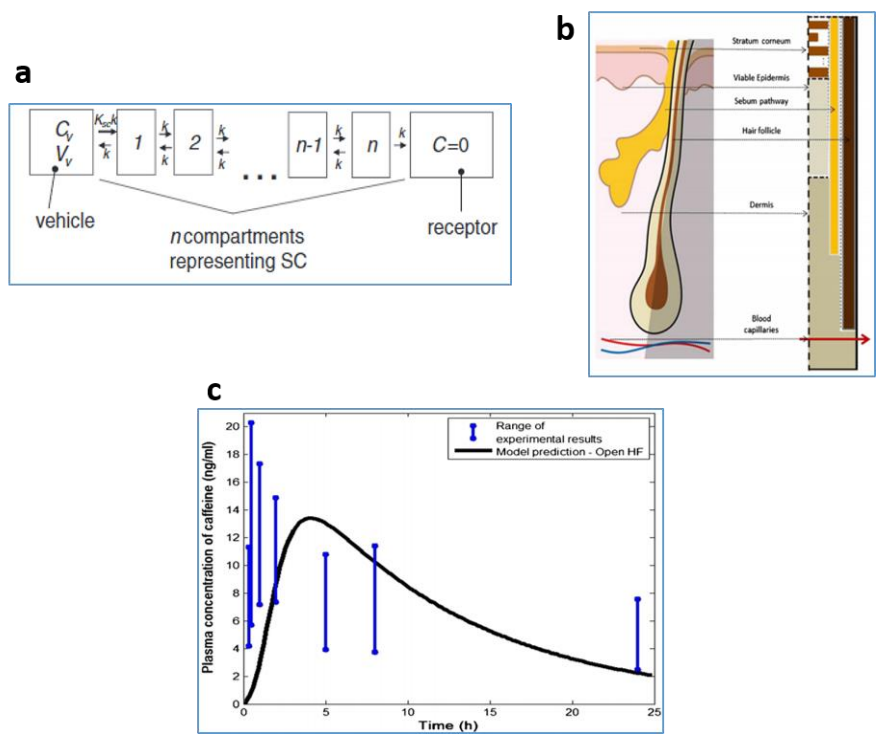

*Figure 2. This Figure intends to highlight the difference between compartmental and DM-PBPK models. a) Schematic of a compartmental model. Reprinted (adapted) with permission from (Anissimov et al. Adv. Drug Deliver Rev. 65(2), pp. 169–190). Copyright (2013) Elsevier; b) Schematic of the microstructure of a hair follicle and its discretized computational domain. Reprinted (adapted) from (Kattou et al. Pharm. Res. 34, pp. 2036–2048). Open Access (2017) Springer Link; c) time-dependent concentration profile obtained by using a FEM approach. Reprinted (adapted) from (Kattou et al. Pharm. Res. 34, pp. 2036–2048). Open Access (2017) Springer Link*

Modelling the hair follicle is particularly important, since it is a weak point through which substances can diffuse rapidly, skipping the keratin barrier in the epidermis and reaching the blood stream. The hair follicle has thus been intensively targeted in recent investigations as the primary route to accelerate drug and cosmetics administration, and as the sole route for big molecules to access the circulatory system [57]. In this case, DM-PBPK models provide more detail relative to QSPR and compartmental models as seen in Figure 2.

There are other specific effects which can be easily considered in a DM-PBPK model. The pharmacokinetics of a xenobiotic substance after skin absorption is influenced by physiological characteristics of the skin, environmental effects and the physico-chemical nature of the compound. The permeability of the skin depends on the location in the body as well. Evidence found in the literature establishes that the absorption rate and final concentration values of a given compound depends on the anatomical site [58], although the assumption of a constant overall permeability coefficient is often considered [59]. Thickness, hydration, and temperature are among the skin-specific factors that affect the microstructure of the different layers. So is the case of the stratum corneum, which presents greater thickness and permeability as the moisture content grows, affecting the mass transfer rate. The change in thickness of the stratum corneum can also be accounted for in DM-PBPK models by using moving boundaries. As for the effect of the environmental factors, Frasch et al. (2014) [60] concluded that wind, humidity, temperature, and vapour pressure affect the quantity of product absorbed by the skin owing to loss of product to the environment through evaporation. To deal with the latter, DM-PBPK models can include mass sink terms at the upper boundary condition to account for the loss of humidity towards the surrounding environment. A frequent source of error when determining xenobiotic absorption rates in *in vitro* cell diffusion studies results from covering the application site with an occlusive wrap to prevent loss of the compound towards the environment. The hydration of the stratum corneum may be perturbed as a result as well, hence affecting the overall absorption rate [61]. Regarding the nature of the xenobiotic compound, its concentration in the environment is usually a factor to be considered, whereas in the case of topical application the vehicle, i.e. the formulation used for topical application, is often considered as an infinite source. Diffusive properties, lipophilicity, polarity, et cetera are also key to determine the final reach of a given compound into the skin [55]. DM-PBPK models provide a high degree of flexibility to account for all the effects mentioned above.

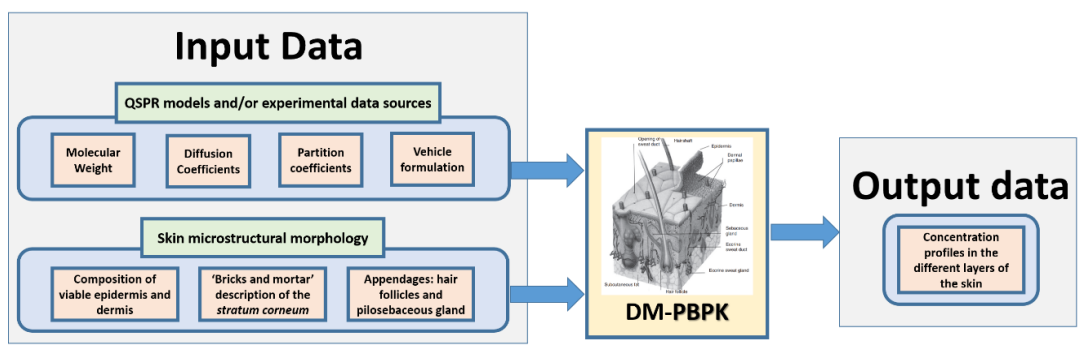

*Figure 3. This input/output schematic diagram shows the data needed to run a DM-PBPK model and the detailed concentration profiles it yields as a result. The main difference with respect to QSPR and compartmental models is that DM-PBPK includes the microstructural morphology of the skin layers. Inset skin microstructure plot reprinted (adapted) with permission from (Anissimov et al. Adv. Drug Deliver Rev. 65(2), pp. 169–190). Copyright (2013) Elsevier*

QSPR models provide however a wealth of data concerning diffusion and partition coefficients which can be used as an input in the development of compartmental and DM-PBPK models (Figure 3). In most cases however, experimental correlations must be used instead. There is always a certain degree of dependency on experimental data and other modelling categories as a result. The combination of all the factors mentioned above results in a complex combination of physical phenomena: evaporation from the surface of the skin, sorption into the stratum corneum, possibility of reversible or irreversible binding, penetration into the viable epidermis, metabolism, and transfer into the blood/lymph stream. Numerical models with ever greater accuracy must be the result of considering more complex features of the skin morphology, i.e. skin appendages, skin layers and vascularization, and the pharmacokinetics of the product at hand, especially metabolism, which can only be achieved by using DM-PBPK models.

The pharmacokinetics of a xenobiotic compound in the organism is divided in four stages: absorption (A), deposition (D), metabolism (M) and excretion (E), which are known in the literature with the acronym ADME. Numerous models have been reported in the literature dealing with absorption, which to date appear to be well validated against experimental data for some specific purposes [62]. While predictive pharmacokinetic absorption models are well established, those dealing with the rest of the aspects of skin pharmacokinetics, namely deposition, metabolism, excretion and others as proposed further down in this document (section 3) are less frequent [63], being notable exceptions the articles published by Jones et al. in 2016 [64] on deposition and by Dancik et al. in 2012 [65] on elimination via the systemic circulation. Among the three least explored aspects of pharmacokinetics, some studies regarding metabolism have been reported lately via QSPR models of data on metabolic routes obtained from the liver (the main organ responsible for metabolism). The skin however is also an important route of exposure, and therefore modelling the skin metabolism is key to predicting and preventing the activity of harmful substances to the human body. Comprehensive and fully predictive *in silico* models of skin metabolism are yet to be developed and would therefore open an important alley in DM-PBPK numerical modelling research.

Pharmacokinetics can therefore be investigated *in vitro*, i.e. using an experimental approach, and/or *in silico*, i.e. computer simulations. Both *in vitro* and *in silico* approaches are complementary. Either because data from experiments have been used to develop the model or to validate predictive models, there must always be feedback between them to develop fully reliable models. Regarding *in vitro*, different approaches exist including empirical testing in bioengineered human tissue specifically grown for that purpose, in *ex vivo* skin samples obtained from living organisms by excision, or in animals (*in vivo*) ([66]; [67]). Numerical modelling on the other hand, reduces experimental costs, and avoids any possible ethical conundrum associated with animal and human testing.

The following subsections provide further detail on the above mentioned *in silico* modelling approaches found in the literature to provide evidence for the improvements proposed in section 4 as future work. The following are further detailed descriptions of the three modelling categories mentioned above.

DM-PBPK models consider the microstructure of the skin and solve the diffusion equation numerically. These models are still in an early stage of development, despite the growing computational power available. QSPR and compartmental models are thus far more utilized for the study of the diffusion of xenobiotics into the skin. Some assumptions must be made when developing a DM-PBPK model given the complexity of the diffusion of xenobiotics into the skin, such as the number of spatial dimensions to be considered. One-dimensional models might be helpful when special microstructural aspects of the skin such as hair follicles or sweat glands are not considered, or when diffusion in the parallel direction to the skin surface is negligible compared to that occurring across the different layers. The difference between one-dimensional DM-PBPK and compartmental models lies thus in considering or not the microstructure of the skin.

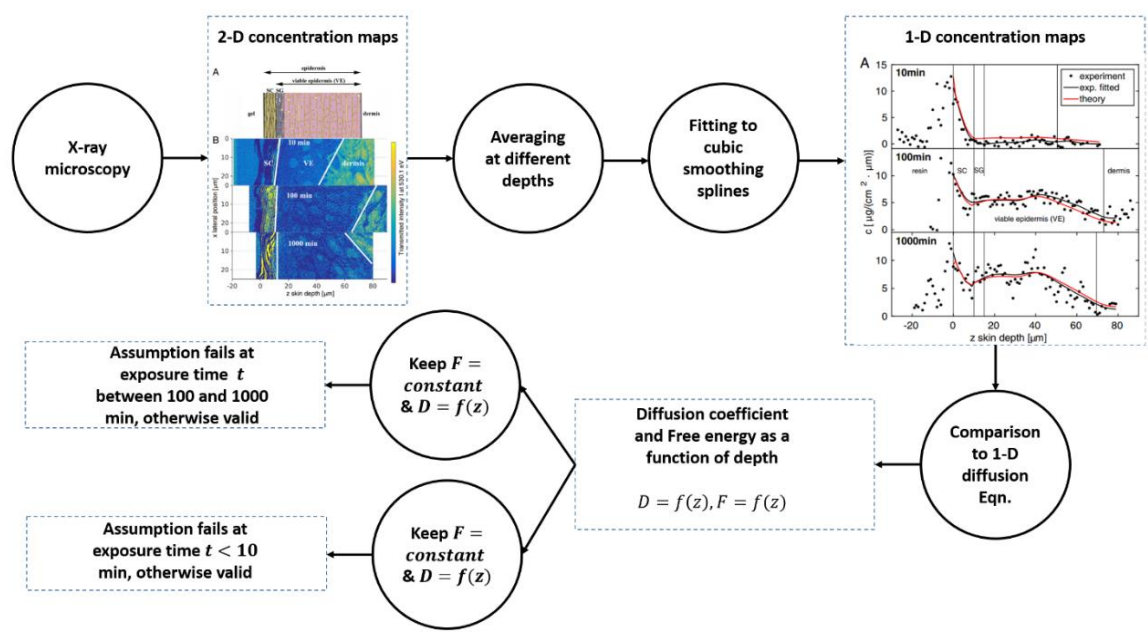

*Figure 4. The schematic shows the information flow of the methodology presented by Schulz et al. (2017). Inset figures have been adapted from Schulz et al. (2017). Inset plots reprinted (adapted) from (Schulz et al. P. Natl. Acad. Sci. 114(14), pp. 3631–3636). 2017 National Academy of Sciences.*

An example of a one-dimensional diffusion-based model was presented by Schultz et al. (2017), the data flow of which has been summarized in Figure 4. The methodology presented by Schulz et al. (2017) [68] can be described as follows. Two-dimensional concentration profiles of the anti-inflammatory drug dexamethasone were obtained using X-ray microscopy. These two-dimensional data were transformed into a one-dimensional profile by averaging at different depths from the skin outer surface and subsequently using cubic smoothing splines to adjust them. The one-dimensional concentration profiles were input to the diffusion equation from which they obtained the diffusivity $D$ and the free-energy $F$ as a function of the depth. The model also relied on experimental work carried out to determine the thickness of the different layers to fix the boundary conditions. Two alternative assumptions were considered:

- Does the model capture the experimental data when considering a constant value of the free energy and a diffusion coefficient varying across the epidermis?
- Does it when $F$ varies and $D$ is constant?

In summary, the authors confirmed the use of the one-dimensional diffusion equation as a valuable method, provided data on the concentration profiles are obtained beforehand. The first assumption failed for exposure times beyond 1,000 min whereas the second failed at earlier times.

The need for an extension to two- or three-dimensional models becomes more evident when further complexity is to be added. Pavlov et al. (2019) [69] presented a numerical study based on the finite-element method (FEM) of transdermal drug delivery in three dimensions in order to account for the shape of the skin pores. The use of a three-dimensional model was in this case necessary given the cross-section shape of the micro-pores. Figure 5 shows a schematic of the pore shape and the space discretization used to study the effect of the pore size on the flux of the xenobiotic and the effect of the vehicle thickness. The authors considered two mathematical expressions to calculate the mass flux crossing the layers of the skin, including the model of Rzhevskiy et al. (2016) [70].

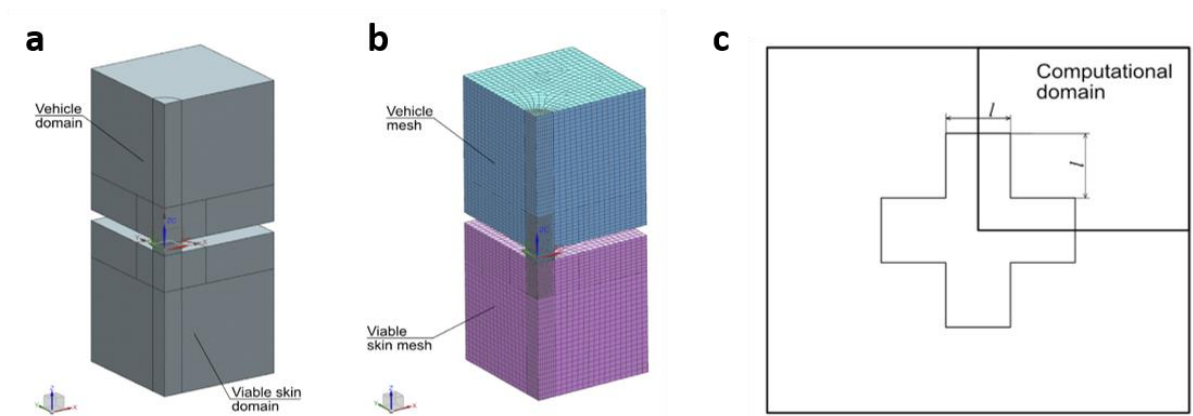

*Figure 5. This figure shows: a schematic of the computational domain (a); the mesh (b); and the shape of the cross-shaped pore seen from above. Reprinted (adapted) with permission from (Pavlov et al. J. Pharm. Sci. 108, pp. 358–363). Copyright (2019) Elsevier.*

The model of Pavlov et al. (2019) illustrates the capabilities of DM-PBPK models versus compartmental ones. The three-dimensional spatial dependence of the concentration is predicted instead of considering separate black box compartments characterized by a constant value of the concentration. The diffusion equation can be solved either analytically based on some assumptions or numerically using finite differences, finite element methods, or finite volume methods.

DM-PBPK models are therefore capable of providing a detailed description of the concentration-depth profile within each skin layer and to consider the effect of the microstructure on the diffusion of xenobiotics across the skin. The values of mass diffusivities and partition coefficients used in DM-PBPK models are most often based on experimental data. *In vitro* assays are thus used to develop and/or validate them. Their capability to be extrapolated to describe different modelling scenarios remains limited and depends on their input and subsequent validation against specific purpose data. DM-PBPK models feature thus the mathematics needed to reach an accurate description of the diffusion process. To gain accuracy, DM-PBPK models need to be complemented with a description of the skin microstructure, such as appendages or blood vessels.

Today, research efforts are directed towards the development of polyvalent models capable of being applied to many scenarios, including metabolic routes and excretion. Data on the validation of DM-PBPK models in the literature are nonetheless scarce. Kattou et al. (2017) [56] reported a comparison between their DM-PBPK model and the experimental data of Otberg et al. (2008), who carried out *in vivo* assays in which a solution with ethanol and propylene glycol containing 2.5% caffeine was topically applied to the chest of six volunteers. The hair follicles were blocked using wax, and sampling was necessary to determine the concentration of caffeine in the blood. In the simulations, only the propylene glycol was considered for simplification. The comparison between *in silico* and *in vitro* data with and without hair follicle blocking shows good agreement, fostering the confidence in this modelling approach. The experimental data show a high degree of variability attributed to the characteristics of the individuals from which data are sampled though, but the order of magnitude is similar in both data series. An offset between the *in vitro* and the *in-silico* concentration peak times (attributed to lack of modelling details concerning metabolism) is however observed in the model of Kattou et al. (2017).

To further increase the accuracy of the models, the next step in its development might well be the inclusion of metabolic routes, i.e. by means of the introduction of mass source terms in the discretized diffusion equations, to represent the generation/consumption of substances derived from the interaction between substances already present in the skin and a given xenobiotic. Moreover, the model presented by Kattou et al. (2017) focuses on both the epidermis (considering the stratum corneum and the viable epidermis) and the dermis, for which the mathematical terms describing convective transport are defined to accommodate the entering of the xenobiotic component to the systemic circulation. The main transport mechanism in the stratum corneum and the viable epidermis is passive diffusion and therefore the convective term is omitted in the mathematical description of these skin layers. The convective term is included in the dermis to account for the presence of blood vessels which can carry the xenobiotic component away. The model can also provide visualization between different possible penetration pathways (opposite to the capabilities of QSPR and compartmental models): the route through the lipids in the stratum corneum and the subsequent viable epidermis and dermis, the intracellular route through the low permeability corneocytes which form the stratum corneum (the structural description of the stratum corneum by means of the bricks and mortar model described the following subsection 2.3.2), and through the hair follicle, which constitutes a shortcut for xenobiotic penetration into the highly vascularized dermis. As boundary conditions, Kattou et al. (2017) considered zero flux outer boundaries, which implies no loss of active component through evaporation to the environment. The method of lines was applied to simplify the governing partial differential equations into linear differential equations [71]. Further simplifications introduced in this model were the suppression of the bending of the stratum corneum in the region nearby the hair follicle and the funnel-shaped infundibulum [72]. Kattou's model was therefore one of the first attempts to solve the diffusion equation on a discretized representation of skin appendages. A similar approach could be easily applied to the study of the sweat gland, for which up to date only microfluidic models have been developed [73].

Different DM-PBPK methodologies were presented by Nagayama and Kurihara (2018) [74] and Safdari and Kim (2019) [75]. The latter applied a Lattice-Boltzmann model to describe the adsorption of drugs into the skin. The former authors proposed instead a particle model following the Lagragian approach to not only describe diffusion, but also represent epidermal formation, trans-epidermal water loss (TEWL) and keratinization of corneocytes.

In summary, thanks to the microstructural description of the skin layers, DM-PBPK models can shed light on further details impossible to achieve using QSPR and compartmental models.

By solving the diffusion equations, detailed concentration maps across the different layers which form the skin can be obtained.

The following subsections will describe in more detail the modelling of the vehicle, stratum corneum, viable epidermis and appendages.

### 3.1. Modelling of the diffusion through the vehicle

In some cases, the vehicle can be approximated as an infinite source of the xenobiotic studied, with a constant concentration as proposed by Chen et al. (2008) [55]. In other practical situations as is the case for slow release patches, a vehicle layer with a finite thickness and active component concentration is more realistic.

Considering the microstructure and the composition of the vehicle are new research directions of interest to industry. These other components might help with improving the delivery, stability or activity of the active ingredient, or simply by facilitating the delivery via an increase in moisture content in the skin. Commonly used components include surfactants and penetration enhancers, which increase the diffusion coefficient of the xenobiotic component by disrupting the barrier which constitutes the stratum corneum. Penetration enhancers may also improve the partitioning between the formulation and the stratum corneum, or decrease the skin thickness to facilitate permeation [76].

### 3.2. Modelling of the stratum corneum: The 'bricks and mortar' approach

The stratum corneum represents the main barrier to xenobiotic penetration and is therefore rate-limiting [77]. The stratum corneum is mainly formed of several consecutive layers of low permeability keratin cells named corneocytes, which result from the replacement of intracellular organelles by a compact proteinaceous cytoskeleton. In the 'bricks and mortar' description, each corneocyte is treated as a block with near-zero permeability linked to the adjacent block through a lipidic gap where diffusion occurs, i.e. the mortar. The composition of the stratum corneum depends on the area of the anatomical location considered though, but on average is formed of 15–25 layers of flattened, hexagonal corneocytes. Each corneocyte has an approximate diameter of 30–50 μm and a thickness which varies between 0.2–0.5 μm [78]. The mortar constitutes between 10% and 15% of the dry mass of the stratum corneum, and therefore, having a high proportion of water, it serves as the main channel for xenobiotic permeation. The implementation of a PBPK model featuring the microstructure of the stratum corneum as a 'bricks and mortar' zone can shed light on the relative importance of those three routes based on different factors such as hydration degree. To model the stratum corneum, the diffusion coefficient of a given component through the corneocytes (intracellular route), and through the mortar (lipidic route), as well as the partition coefficients between the lipids and the corneocytes, should be determined. Anisotropic diffusion in the lipids of the stratum corneum, which is caused by its heterogeneous composition (crystalline and liquid parts), is usually neglected in the models reported in the literature. Bouwstra et al. (2002) [79] considered the lipid phase in the 'bricks and mortar' model as homogeneous, resulting in an approximation which has proved to be accurate enough. Detailed description of its microstructure and the three possible pathways through it (intracellular route through low permeability corneocytes, high permeability lipidic route, and appendageal shortcut) can reduce the gap between *in vitro* and *in silico* data. Recently developed diffusive-based models have incorporated a detailed description of the morphology of the stratum corneum known as the 'brick and mortar' model [80].

Naegel et al. (2009) [81] and D. Feuchter et al. (2006) [82] developed a three-dimensional computational model to account for the effect of the corneocyte shape on the overall diffusion of xenobiotic components through the stratum corneum. They implemented corneocytes with—in increasing order of realism—ribbon, cuboid, and tetrakaidekahedral shapes based on previously reported literature data ([83]; [84]), reaching the conclusion that the morphology of the corneocytes plays a key role on the permeability of the stratum corneum. A tetrakaidekahedral corneocyte resulted in skin permeability that doubles that found for 2-D ribbons. Their work has application in the understanding of skin diseases such as psoriasis, which affect the morphology of the stratum corneum, and thus, results in a change of skin permeability. The treatment of the stratum corneum as a heterogeneous layer due to the existence of both corneocytes and the lipidic gap implies a considerable computational effort. Muha et al. (2011) [85] developed in this regard a mathematical method to implement a homogeneous equivalent to the brick-and-mortar model. Such simplifications may be taken when there is no need to quantify the difference between the three routes across the stratum corneum.

In order to develop a comprehensive model of the stratum corneum, the defining geometric values and composition of the stratum corneum are provided by Johnson et al. (1997) [86] and Mitragoti et al. (2003). Apart from the location in the body considered, the thickness of the stratum corneum presents a strong dependency on the hydration degree, varying from 52 μm to 16 μm upon low hydration, which affects the diffusion of any xenobiotic component strongly. The effect of hydration on the change of thickness and permeability of the stratum corneum can be studied by way of using moving boundaries (a capability available in commercial finite element method software).

### 3.3. Modelling of the hair follicle, the viable epidermis and the dermis

Modelling of the appendages in skin can include the hair follicle and the pilo-sebaceous unit, although the former has attracted a greater share of research effort in the last few years. The hair follicle can act as a shortcut for xenobiotic penetration when sebum is being produced and the hair follicle is growing (Krause and Foitzik, 2006 [87]). The hair follicle has not been traditionally considered as a penetration route because of the small proportion of area they occupy relative to the total surface of the skin [88]. Hair follicles possess high vascularization however which could turn them in shortcuts for xenobiotic administration. They also provide higher area than initially thought because they reach deep into the dermis, hence increasing the actual area available for xenobiotic permeation. Hair follicles present also diffusion coefficients significantly higher than the stratum corneum, making the follicular route more interesting from the point of view of xenobiotic absorption than the rest of the skin surface [89]. Some studies consequently suggest that the follicular route is particularly favourable for hydrophilic and high molecular weight molecules, as well as in the case of drugs delivered by using particles, owing to the difficulty of these molecules to cross the stratum corneum [90]. In spite of those advantages, models of the follicular absorption route are scarce in the literature, being the reported models presented by Liu et al. (2011) [91], Bookout et al. (1997) [92], and Frum et al. (2006) [93] summarised in Table 3. Liu et al. (2011) [91] compared the penetration of caffeine in human skin blocking and without blocking hair follicles using two different methods including a compartmental model with first order absorption and elimination. Bookout et al. (1997) also reported a compartmental model including parallel sub-compartments to represent the hair follicles. An interesting modelling approach is that of Radtke et al. (2017) [94] who presented an approach based on Brownian dynamics instead of solving diffusion equations. To the best of the authors' knowledge, the only DM-PBPK model of the hair follicle is that of Kattou et al. (2017) [56]. The models based on Brownian dynamics, i.e. molecular dynamics, and the DM-PBPK model provide a better visualization of the penetration of a xenobiotic into the skin. Figure 6 shows a snapshot of both Radtke's computational domain and results. Their model was used to assess the effect of radial and axial movement on the advancement of nanoparticles towards the dermis. On the other hand, the compartmental model of Liu et al. (2011) proves as an adequate approach in order to assess the effect of the presence of the hair follicle in the system as a whole and obtain the time-dependent concentration of the xenobiotic in plasma, but lacks the detail at the smaller scale that Radtke (2017) and Kattou's model (2017) are able to provide.

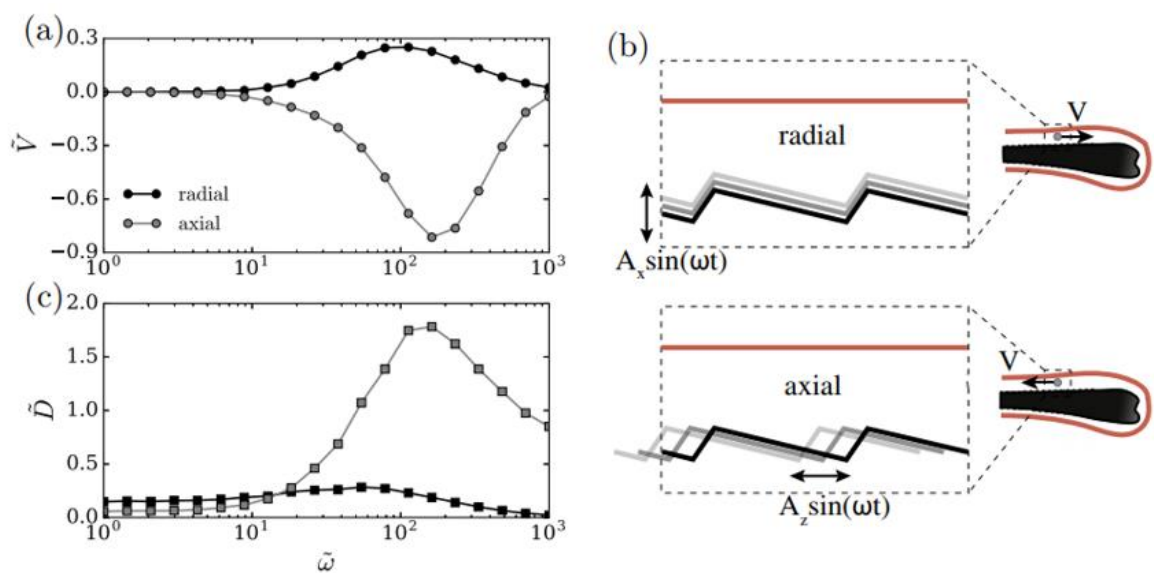

*Figure 6. Velocity of the nanoparticle along the axis of the hair follicle as a function of the frequency of the hair follicle movement (a); schematic of radial vibration of the hair follicle and the forward movement of the nanoparticle (b); diffusivity of the nanoparticles as a function of frequency (c); and schematic of the axial vibration of the hair follicle. Reprinted (adapted) with permission from (Radtke et al. Eur. J. Pharm. Biopharm. 116, pp. 125–130). Copyright (2017) Elsevier.*

*Table 3 Summary of relevant literature on hair follicle modelling*

| Reference | Method | Conclusions |
|---|---|---|
| Liu et al. (2011) [91] | Wagner-Nelson method and compartmental with first order absorption and elimination. | - Tenfold increase of absorption rate through follicles with respect to absorption rate through stratum corneum.<br>- No absorption delay in follicles; whilst 10 min delay in the stratum corneum. |
| Bookout et al. (1997) [92] | Enhanced PBPK model obtained with measurable parameters and featuring parallel layers representing the skin appendages | - Improved predictions of chemical concentrations are obtained with the multi-layered model rather than with |
| | | the previous homogeneous model in which the present model is based. |
| Frum et al. (2006) [93] | Skin sandwich technique. | • Above a critical value of octanol-water partition coefficient ($K_{ow}$), the relation between the % follicular contribution and the drug absorption rate depends on the lipophilicity. |
| Radtke et al. (2017) [94] | Brownian dynamics | • Penetration enhancement through hair follicles explained by the ratchet mechanism. |
| Kattou et al. (2017) [56] | Predictive model using open code. | • Data with which the model is compared is not used to predict the diffusion parameters which define the model. |

As far as the viable epidermis and the dermis are concerned, both behave similarly given their similar compositions. The dermis consists of collagen and elastin fibres immersed in an aqueous matrix, with a thickness which depends on the anatomical location and reaches 4 mm at its maximum. The hair follicles and sweat glands are embedded within the dermal layer, and thus provide a fast route towards the rich vascularisation located in the transition between the epidermis and the dermis [95]. The presence of drug-binding proteins and small ramifications of the blood and lymphatic system is also worth considering. Dermic blood vessels are concentrated in the upper 100–200 μm of the dermis, i.e. the papillary dermis, and decrease considerably in the lower part of the dermis, i.e. the reticular dermis. These aspects are also important to define areas of the computational domain where convection should be considered. The overall structure of the vessels consists of a dense aggregate of blood vessels running along the epidermal–dermal junction, a less dense plexus found much deeper in the lower (reticular) dermis, and vessels that also connect to the hair follicles and sweat ducts [96]. The viable epidermis and dermis are therefore straightforward areas to be modelled, given their rather homogeneous nature. Two possible further modelling steps consist in including the convective transport in blood vessels, and the drug-to-protein binding, which can be modelled as negative mass sources in the diffusion equation.

## 4. Conclusions and perspectives

This review gives an overview of physiologically-based pharmacokinetic (PBPK) models to predict the diffusion of xenobiotics into the skin and focuses on the discretized-microstructure (DM) physiologically-based pharmacokinetic (PBPK) models.

QSPR models—also known in the literature as QSAR models—are the result of phenomenological fittings. QSPR models offer an overall estimation of the absorption and thus, they constitute useful tools for risk assessment purposes. These models lack however a detailed physical description of the skin morphology. QSPR models can be used as a basis to develop more complex models.

Compartmental models describe the interaction between the different skin layers and skin appendages but lack some detail on what occurs inside those skin layers. To reduce the calculation time, compartmental models introduce and solve linearized diffusion equations, providing a single representative value of the concentration in each compartment. Compartmental models provide thus an estimation of the concentration in each skin layer, as well as the transfer rate between compartments. With respect to QSPR, compartmental models represent a first estimation of the actual distribution of xenobiotics concentration within the skin, but lack detail since they neglect the morphology of the skin layers.

To bridge this gap, Discretized-Microstructure (DM) PBPK models have been introduced, to provide the user with detailed information on the concentration profiles across the skin without having to recur to overdetailed and computationally-expensive molecular dynamics simulations. These models can consider the microstructure of the stratum corneum (via the 'bricks and mortar' approach), and other routes such as the hair follicle route, which is of high interest to the cosmetic industry and for drug delivery purposes since it has been experimentally proved as an effective shortcut to reach the highly vascularised dermis. DM-PBPK models also allow for insight on the different penetration routes (low permeability intracellular route, high permeability lipidic route, and appendageal route), establishing the relative importance between them as a function of the physico-chemical nature of the particular compound studied, mainly the size of the molecule, its lipophilicity, and its solubility in water. Despite the increased degree of microstructural detail, the capabilities of DM-PBPK models have not been widely exploited yet and further validation is needed.

Some promising research directions to expand the capabilities of DM-PBPK models are (Figure 7):

a) Considering the effect of vehicle thickness, microstructure and physical properties on xenobiotic transfer.

The capabilities of DM-PBPK models could be relatively easily expanded by considering the microstructure of the vehicle and the slow release of active principles or the evolution of the vehicle properties while its solvent evaporates or diffuses into the skin. Furthermore, the transfer of nano-particles (e.g. $TiO_2$ and $ZnO_2$ in sunscreens) through the skin to the blood stream, which is a potential major side effect of their use in cosmetics ([97]; [98]; [99]), could be considered. This could support the development of novel cosmetic formulations able to slow down or to stop the transfer of nanoparticle to the skin and prevent/limit their negative effects on the skin and on the body.

b) Considering the effect of skin hydration on the transfer of xenobiotics and on the skin microstructure

Skin hydration causes swelling of the corneocytes, changing the microstructure of the stratum corneum and its permeability to the active principles found in a given formulation. Experimental evidence suggests that the swelling of the stratum corneum—the main barrier to xenobiotic penetration—is mainly due to the change in shape of the corneocytes, particularly in the direction perpendicular to the skin surface, resulting in a reduction of their aspect ratio. Bouwstra et al. (2003) [100] provide quantification of this phenomenon and suggest that not all the corneocytes swell to the same extent, with those near the outermost part of the stratum corneum showing less aspect ratio than those at deeper parts. The change in microstructure and composition of the stratum corneum caused by hydration results in the modification of its permeability. Owing to the moisture increase causing a reduction of the proportion of low-permeability keratine in the stratum corneum, there is an exponential increase in the diffusion coefficient with the moisture content. There is thus a combined effect of moisturization, with an increase of thickness of the stratum corneum on the one hand and an increase in the diffusion coefficient on the other. The former should result in hindered xenobiotic penetration owing to the increase in the thickness of the main barrier to penetration, whereas the latter should enhance it. The solving of the diffusion equations in a DM-PBPK model representing the microstructure of the skin should constitute a powerful tool for the analysis of this contrasting effects. The swelling/shrinking of the stratum corneum can be accounted for by using moving boundaries in a finite element method model. On the other hand, the change in diffusivity owed to swelling/shrinking can be accounted for by introducing a thickness dependent value of the diffusivity in the simulation set-up.

c) Modelling perspiration and its effects on xenobiotic transfer

Skin perspiration consists in the secretion of sweat by the sweat glands, its transport through the pores toward the skin surface and the formation of a liquid layer at the surface of the skin. The effect of sweat secretion on the skin moisture content, on the delivery of actives and on the undesired influx of nanoparticles should be evaluated. If the sweat duct can act as a preferential path for xenobiotic ingress, then the effect of the outflow of sweat should play an important role in modulating the xenobiotic transfer into the skin. The modelling of this effect can be easily accomplished by adding mass source terms in the diffusion equations, defined only in the relevant computational cells.

   d)  Introduction of metabolic pathways

Metabolic reactions of the compounds absorbed in the human skin can also be modelled by adding mass sink terms to the governing diffusion equation. Those terms depend on the reaction rate and account for the generation/consumption of components owing to chemical reactions. The addition of metabolic routes and chemical binding will potentially reduce the gap between experimental and numerical data [63]. Cutaneous metabolism of a chemical applied to the skin is difficult to differentiate *in vivo* from the systemic metabolism occurring mainly in the liver by analysing blood and excreta samples. *In vitro* approaches can provide accurate insight on the specifics of skin metabolism by isolating the skin from the metabolic activity in the rest of the body, providing a source of data for validation purposes.

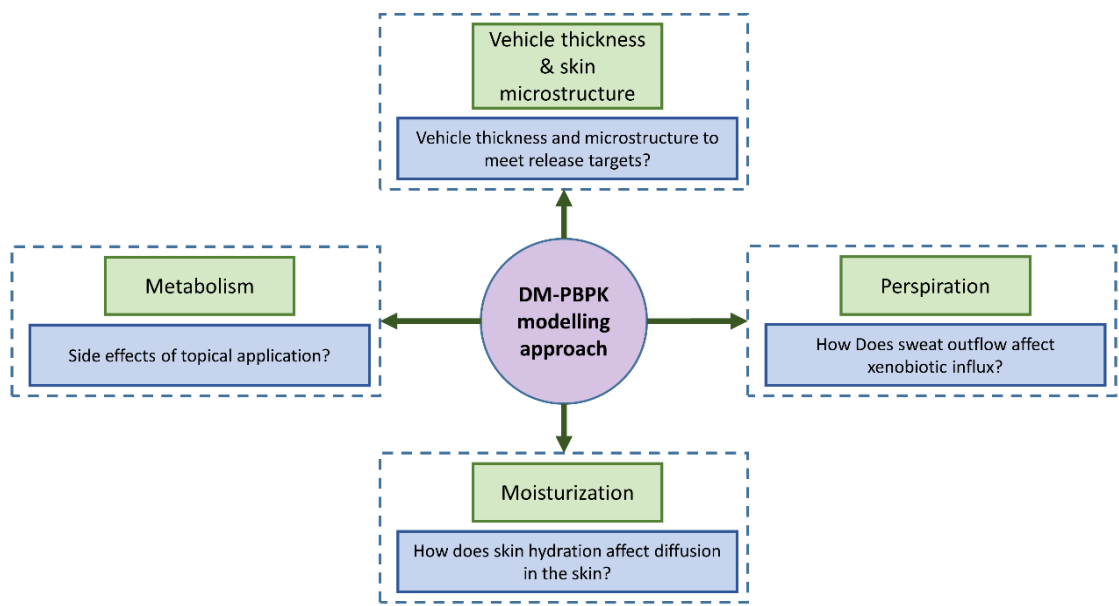

*Figure 7 Schematic view of future research directions and capabilities of the diffusion-based PBPK modelling approach*

**Acknowledgements**

The authors would like to express their gratitude to Nestlé Research Center in Vevey (Switzerland) for the funding provided to carry out this review.